\documentclass{article}
\usepackage{epsfig,graphicx}
\usepackage{float}
\usepackage{amsmath}
\usepackage{authblk}

\newcommand{\be}{\begin{equation}}
\newcommand{\ee}{\end{equation}}
\newcommand{\cotan}{\mathrm{cotan}}

\newtheorem{theorem}{Theorem}[section]
\newtheorem{definition}[theorem]{Definition}

\title{Spectral solution of load flow equations}

\author[1]{\normalsize{J. G. Caputo }\thanks{caputo@insa-rouen.fr}}
\author[1]{A. Knippel \thanks{arnaud.knippel@insa-rouen.fr}}

\affil[1]{Laboratoire de Math\'ematiques, INSA de Rouen Normandie\\ 76801 Saint-Etienne du Rouvray, France.}

\author[2]{\normalsize{N. Retiere} \thanks{nicolas.retiere@univ-grenoble-alpes.fr}}
\affil[2]{Grenoble Electrical Engineering Laboratory (G2Elab),
University of Grenoble Alpes (UGA), CNRS, F-38000 Grenoble,}

\date{ }

\begin{document}
\maketitle
\vspace{-1cm}

\date{\ }

\begin{abstract}
The load-flow equations are the main tool to operate and plan
electrical networks. For transmission or distribution networks
these equations can be simplified into a linear system involving
the graph Laplacian and the power input vector. 
Decomposing the power input vector on the basis of the
eigenvectors of the graph Laplacian, we solve this singular
linear system. This spectral approach gives a new geometric view 
of the network and power vector. The power in the lines is then given 
as a sum of terms depending on each eigenvalue and eigenvector. We
analyze the effects of these two components and show the important
role played by localized eigenvectors. 
This spectral formulation yields a Parseval-like relation for
the $L_2$ norm of the power in the lines. Using this relation
as a guide, we propose to consider only the first few eigenvectors
to approximate the power in the lines. Numerical results on IEEE cases
support this approach which was also validated by analyzing
chains and grids.
\end{abstract}

\maketitle

\section{Introduction}

The electrical grid is one of the major engineering achievements
of the 20th century. Typically, it involves high voltage transmission
lines connecting large generators and power sub-stations. The 
distribution network starts at the sub-station and delivers the 
energy to the end user.
The grid was originally designed to distribute 
electricity from large generators. It is changing rapidly due to the
emergence of renewable and intermittent sources, energy 
storage and electric vehicles \cite{bc13}. Not addressed properly, this
complexity could result in management difficulties and possible
black-outs. An important issue for the network is to predict 
the power in the lines and identify critical lines, {i.e.} 
the ones that are most heavily loaded. 
The network should be planned and operated to control the load on 
these lines.

The main model used by operators and planners to analyze stationary electrical
networks is the so-called load-flow equations\cite{kundur},
connecting incoming
power to voltage and current. These equations are nonlinear.
Typically they are solved using a Newton method \cite{gsc15}. They can
have multiple solutions and the iteration scheme can fail to converge.
In any case, it is difficult 
to see how the load-flow solution is affected by the topology of 
the network and the nodal distribution of generators and loads.
A global geometrical
point of view, incorporating the topology and the load-generator 
distribution would be very useful to address this issue.

In this article, we propose such a geometrical point of view.
We consider the case of a transmission network and 
linearize the load-flow equations to obtain a 
Laplacian equation, involving
the graph Laplacian operator associated to the network \cite{crs01}.
This matrix can be seen as a discrete version of the continuous
Laplacian, see for example the finite difference approximation
in numerical analysis, see for example \cite{ananum}. 
The Laplacian matrix is positive and symmetric so that its eigenvectors
can be chosen orthonormal. Using these eigenvectors, we introduce 
a spectral solution of the Laplace equation.
This is a Fourier like picture of the network, where the small
order eigenvectors correspond to large scale fluxes on the network.
Conversely large order eigenvectors correspond to small scale
fluxes on the grid. 
Our spectral picture naturally shows the
dependence of the line power fluxes on the topology and load-generator  
distribution. 
Similar ideas can be developed for distribution networks, however
these usually have a simple tree like geometry. Then the 
network topology plays a less important role. Also large
scale failures occur on transmission networks. We therefore concentrate
on these networks.

Using the solution of the Laplace equation, we can write
explicitly the vector of power fluxes $P_l$ using the
discrete gradient, $\nabla $ {i.e.} the transpose of the incidence matrix
of graph theory \cite{crs01}. The vector $P_l$ can then be written
as a sum of terms $\nabla \mathbf{v}^i / \omega_i^2$ where 
$\omega_i^2$ is an eigenvalue of the Laplacian with eigenvector 
$\mathbf{v}^i$. We analyze how these two terms affect $P_l$ by examining
the evolution of $\omega_i^2$ and $\mathbf{v}^i$ with $i$. 
We obtain an explicit Parseval relation for $\parallel P_l \parallel_2^2$ 
which can be used for minimization. This shows that to minimize 
the norm of $P_l$ it is crucial to control its components on the
low order eigenvalues, in other words the large scales of the network. The 
role of the eigenvector structure is more difficult to understand, we
therefore examine situations where the generator/load vector is
concentrated on a single $\mathbf{v}^i$. This reveals the importance of
localized eigenvectors that contribute strongly to 
$\parallel \nabla \mathbf{v}^i \parallel$.
Finally, examining more realistic generator/load distributions
shows that truncating the sum for $P_l$ gives a reasonable estimation
so that the full modal decomposition is not necessary. \\
The article is organized as follows. Section 2 recalls the load-flow equations
and how they can be approximated by a Laplace equation. 
We introduce the spectral solution of this equation in section 3.
Section 4 presents the spectrum of some IEEE networks and section 5
illustrates the spectral solution of the reduced load-flow. Conclusions
are presented in section 6.

\section{The load-flow equations }

To introduce these equations, we will follow the very clear
derivation of \cite{panciatici}.
At each node, we write conservation of power, this means :
\be \label{power}
{\cal P }= {\cal V } {\cal I }^* ,\ee
where ${\cal P }$ is the vector of powers inserted into or extracted 
from the network, each component corresponding to a node. The right
hand side is the power due to the network.
From the generalization of Ohm's law 
\be \label{ohm}
{\cal I }= (G + j B) (V +j W) ,\ee
where $G + j B$ is the so-called $Ybus$ matrix \cite{kundur}. We then get
\be \label{ohmc}
{\cal I }^* = (G V - B W ) + j( -B V - G W) . \ee
Combining (\ref{power}) and (\ref{ohmc}), we obtain
\be \label{powerf}
{\cal P }= V (G V - B W ) + W ( B V + G W) +j [ W(G V - B W ) + V( -B V - G W)] .  \ee
Introducing the vector of active and reactive powers, so that
\be \label{act_react}
{\cal P }= P +j Q , \ee
we obtain our final load flow equations \cite{panciatici} :
\begin{eqnarray} \label{lflow}
V (G V - B W ) + W ( B V + G W) = P, \\
W (G V - B W ) + V( -B V - G W) =Q. 
\end{eqnarray} 
In index notation, the system reads, for all nodes $k$ 
\begin{eqnarray} \label{lflowi}
V_k \sum_i (G_{k i} V_i - B_{k i} W_i ) + W_k  \sum_i ( B_{k i} V_k 
+ G_{k i} W_i) = P_k, \\
W_k \sum_i (G_{k i} V_i - B_{k i} W_i ) - V_k  \sum_i (B_{k i} V_k 
+ G_{k i} W_i) = Q_k .
\end{eqnarray} 
The sums correspond to matrix-vector multiplications while
the terms on the left of the sums correspond to tensor products. The two
operations do not commute. The system (\ref{lflow}) is 
quadratic in $V$ and $W$ and needs to be solved using
an optimization solver, for example a Newton-Raphson method.

An important fact is that
the matrices $B$ and $G$ are graph Laplacians \cite{crs01}.
Typical approximations can
be done for the transmission network and for the distribution
network. We examine these in the next section taking advantage
of the special property of $B$ and $G$.

\subsection{Simplified model of a transmission network}

For a transmission network, we follow the three classical assumptions,
(see Kundur's book for example \cite{kundur}):
\begin{itemize}
\item neglect the ohmic part of the $Ybus$ matrix so take $G=0$
\item assume that voltage modulus is constant and close to 1
\item assume that the phase is small
\end{itemize}

Taking $G=0$ leads to the new system
\begin{eqnarray} \label{lflowr}
-V ( B W ) + W ( B V) = P, \\
-W ( B W ) - V( B V) =Q. 
\end{eqnarray}
The second assumption and third assumptions imply 
\be \label{volt_pha}
{\cal V }= V+ jW \equiv v e^{j \theta}\approx 1 + j  \theta ,\ee
because the vector $v \approx 1$. Then the vectors $V,W$ are 
$$V=1 , ~~W =\theta  .$$
The first equation of (\ref{lflowr}) reduces to
\be \label{bteta}
-B \theta = P.\ee
This is a singular linear system to be solved for 
the vector of phases $\theta$ 
knowing the vector of active powers $P$.
To identify critical links we compute the power line vector $P_l$
whose components are the powers in each line. It is calculated using
the discrete gradient $\nabla$ (see \cite{cks13} for an example)
\be \label{pl}
\nabla \theta = P_l .\ee
Note also the connection between $\nabla$ and the graph
Laplacian $B\equiv \Delta = \nabla^T \nabla$.
The two equations (\ref{bteta}-\ref{pl}) are the main 
model that we will consider in the rest of the article. Since the 
matrix $B$ is a graph Laplacian, it is singular and the linear
system (\ref{bteta}) needs to be solved with care. In the next section, 
we will use the important symmetries of $B$ to solve (\ref{bteta}).

In the following we consider for simplicity that all lines
are the same. Then the discrete gradient
$\nabla$ has entries $\pm 1,0$. The Laplacian elements $\Delta_{ij}$ 
are such that $\Delta_{ij}=1$ if node $i$ is connected to node $j$
and $\Delta_{ii} = -\sum_{j\neq i} \Delta_{ij}$, the number of 
links (degree) of node $i$ \cite{crs01}.
This simplification is for clarity of exposition. The whole of our 
spectral formalism presented below carries through 
when the lines are unequal i.e. in the presence of weights.

To conclude this section, note that for distribution networks, a similar simplification of the load-flow equations can be done \cite{kundur}. For those networks, we can assume $$ B=0,~~W \approx 0,~~V = 1 + \delta V . $$ This leads to the following equation, very similar to (\ref{bteta}) \be\label{reduc_dist} P = G \delta V . \ee
In the rest of the article, we will focus on transmission networks.

\section{Spectral solution of the reduced load-flow}

In this section, we use the notation from graph theory and
note the graph Laplacian matrix $B$, $\Delta$.
The matrix $\Delta$ is symmetric and positive.
Its eigenvalues can be written
$$\omega_1^2=0  \le \omega_2^2 \le \dots \le \omega_n^2 ,$$
where $n$ is the number of nodes of the network. The eigenvectors
$$\mathbf{v}^1,\mathbf{v}^2, \dots \mathbf{v}^n  ,$$
can be chosen orthonormal. In the rest of the article, we assume that
the network is connected so that $\omega_1^2=0 < \omega_2^2$ 
\cite{crs01}.

A standard way of solving equation (\ref{bteta}) is to
use the Penrose pseudo-inverse with a regularization \cite{numrec}
to eliminate the singularity due to the zero eigenvalue.
This does not give much information on the way the solution
depends on the graph and the power distribution. To gain
insight, it is useful to project $P$ on the eigenvectors 
\be\label{proj_p}
P= p_1 \mathbf{v}^1 + p_2 \mathbf{v}^2+ \dots + p_n \mathbf{v}^n ,
\ee
and take advantage on their orthogonality.
Assuming that demand and supply are balanced in the electrical generation,
the power vector $P$ satisfies
$$\sum_{k=1}^n P_k =0, $$
where the $P_k$'s are the component in the canonical basis.
Using the expansion \eqref{proj_p}, we get
$$\sum_{k=1}^n P_k = \sum_{i=1}^n p_i (\sum_{k=1}^n \mathbf{v}^i_k) 
= {p_1  \over \sqrt{n}}=0,$$
because the eigenvectors $\mathbf{v}^i$ satisfy $\sum_{k=1}^n \mathbf{v}^i_k=0, i>1$.
Then we get $p_1=0$. One can then calculate $\theta$ as
\be\label{teta}
\theta = -{ p_2 \over \omega_2^2} \mathbf{v}^2  
-{ p_3 \over \omega_3^2} \mathbf{v}^3 \dots -{ p_n \over \omega_n^2} \mathbf{v}^n .\ee
The power in the lines $P_l$ is then
\be\label{plt}
P_l = \nabla \theta = -{ p_2 \over \omega_2^2} \nabla \mathbf{v}^2
-{ p_3 \over \omega_3^2} \nabla \mathbf{v}^3 \dots 
-{ p_n \over \omega_n^2} \nabla \mathbf{v}^n .\ee

Let us now be specific about the distribution of generators and
loads in the network. We introduce the vectors $G$ and $L$
and their components
\be\label{gl}
G = \sum_{i=1}^n g_i \mathbf{v}^i,~~L = \sum_{i=1}^n l_i \mathbf{v}^i, ~~P \equiv G -L . \ee
The euclidian norm of $P_l$ has a particularly simple form. To see this
we write  
$$ \parallel P_l \parallel ^2_2 
= \sum_{i,j=2}^n {(g_i-l_i)(g_j-l_j) \over \omega_i^2 \omega_j^2}
(\nabla \mathbf{v}^i)^T \nabla \mathbf{v}^j.$$
Note that
$$ (\nabla \mathbf{v}^i)^T \nabla \mathbf{v}^j = (\mathbf{v}^i)^T (\nabla^T \nabla) \mathbf{v}^j
= (\mathbf{v}^i)^T \Delta \mathbf{v}^j = \omega_i^2 \delta_{ij} ,$$
where $\delta_{ij}$ is the Kronecker symbol.
We get finally the Parseval like relation
\be\label{pl2}
 \parallel P_l \parallel ^2_2 
= \sum_{i=2}^n {(g_i-l_i)^2 \over \omega_i^2 } .
\ee
This simple expression shows that the $L_2$ norm of the power depends
only on the eigenvalues and the projections of the input-output powers
on the eigenvectors. In the following we will use this expression
to guide the changes to the generator or load distributions. 
Expression (\ref{pl2}) also holds for the weighted Laplacian, so that
(\ref{pl2}) can be used for real electrical networks.

\subsection{Theoretical background : nodal domains}

The eigenvectors $\mathbf{v}^i$ give rise to the so-called nodal
domains. We recall the following definitions and theorem 
following the presentation of \cite{booklapla}.
\begin{definition}[{\rm Nodal domain }]
\label{ndomain}
A positive (negative) nodal domain of a function $f$ defined on
the vertices of a graph $G(V,E)$ is a maximal connected 
induced subgraph of $G$ on vertices $v \in V$ with $f(v) \ge 0$ 
($f(v) \le 0$).
\end{definition}
For a strong positive nodal domain, the sign $\ge$ should be replaced by $>$.
In the electrical grid context, positive nodal domains correspond to 
generators while negative nodal domains are loads.

We call ${\cal S}(f), {\cal W}(f)$ , respectively the positive strong and weak 
nodal domains of a eigenfunction $f$. 
We have the following result \cite{gladwell}.
\begin{theorem}[{\rm Discrete nodal domain theorem}]
\label{bound_ndomain} Let $\Delta$ be a generalized Laplacian 
of a connected graph with $n$ vertices. Then, any eigenfunction $f_k$ 
corresponding to the $k$th eigenvalue $\lambda_k$ with multiplicity 
$r$ has at most $k$ weak nodal domains and $k+r-1$ strong nodal domains.\\
${\cal S}(f_k)  \le k,~~~~~{\cal W}(f_k)  \le k + r - 1$.
\end{theorem}
Then, the nodal domains are small (resp. large) scale for 
large (resp. small) $i$. In particular, the eigenvector corresponding to
the first non zero eigenvalue partitions the graph in two sub-graphs, see
the following result from Fiedler \cite{fiedler}.
\begin{theorem}
An eigenfunction of second eigenvalue has exactly two nodal domains.
\end{theorem}

The power in the lines $P_l$ is connected to
the vectors $\nabla \mathbf{v}^i$. These in turn, depend on the nodal
domains. We see in the next section, how eigenvectors $\mathbf{v}^i$ that
have small nodal domains will have large $||\nabla \mathbf{v}^i||$ which will
contribute strongly to $||P_l||$.

\subsection{Decay of inverse of eigenvalues}

We have the following inequality \cite{mohar91} for $\omega_2^2$
\be\label{mohar} 
{4 \over n D } \le \omega_2^2 \le {n \over n-1} , \ee
where $D$ is the diameter of the graph, {i.e.} the maximum distance
between two vertices.
We denote by $deg(u)$, the degree of vertex $u$, {i.e.}: the number
of edges incident to $u$.
The maximal eigenvalue is such that \cite{mohar91}
\be\label{mohar2} \omega_n^2  \le { \rm max } \{ deg(u)+deg(v),~~ uv ~~{ \rm edge 
~of~ G } \} . \ee

Typically electrical networks have an average degree $2 \le  {\bar d} \le 3 $.
Assuming that the maximal degree is bounded, then $\omega_n^2$ 
will be bounded from above as $n$ increases. Take for example
a grid, the inequality reads $\omega_n^2 \le 8$; in fact
$\omega_n^2=8$ so the inequality is sharp. 
On the other hand, the lower bound ${4 \over n D }$ of $\omega_2^2$ 
decreases as $n$ increases. We then expect the spectrum of the Laplacian
to extend more towards $0$ as the network gets larger.

\subsection{Practical consequences for electrical networks}

The spectral approach that we present gives a geometric
picture of the network and the power vector. It gives
a quick approximation of the solution of the nonlinear load-flow
equations.

Relation (\ref{pl2}) gives explicitly the L$_2$ norm of the
energy in the lines. This remarkable result provides a way to optimize
the electrical network. 
The relation (\ref{pl2}) implies that taking $g_i=l_i$ makes the power in
all the lines zero. This corresponds to not having any network.
Each node has a generator exactly balancing its load. This is of
course not reasonable. Instead (\ref{pl2}) seems to indicate that
the dominating terms are the small $i=2,3,4..$ terms.
Then, to minimize the expression we can choose the corresponding amplitudes 
$g_i-l_i$  to be small. This naive analysis will be checked carefully
and confirmed below.

If the infinite norm is required, then
we just use formula \eqref{plt}. 
The following bounds for the L$_\infty$ norm can be used
\be\label{linf2}
{1\over \sqrt{n}} { \parallel P_l \parallel _2 } \le  \parallel P_l \parallel _\infty  \le  \parallel P_l \parallel _2 .
\ee
The infinite norm will provide the line carrying the most power, i.e. 
the most critical line.

\section{Spectral features of some IEEE networks}

In this section, to estimate the relative influence of $\omega_i^2$ 
and $\nabla \mathbf{v}^i$, we input the power on a single eigenvector, 
$$ P= p_i \mathbf{v}^i, ~~~ 2 \le i\le n , $$
and $p_1=g_1-l_1=0$.
To be able to compare different $i$,
we choose $p_i$ so that the sum of the
positive components is equal to 1, this corresponds to having
an equal generator (or load) power in the network independently of $i$.
It is equivalent to setting $\parallel P \parallel_1=2$.
We examine two IEEE networks, with 30 and 118
nodes respectively and use the parameters given in the files of
the Matpower software \cite{matpower}.
The loads are chosen uniform on the network, i.e. $l_1=g_1$
and $l_i=0,~~i\ge 2$.

\subsection{IEEE Case 30}

\begin{figure}[H]
\centerline{ \epsfig{file=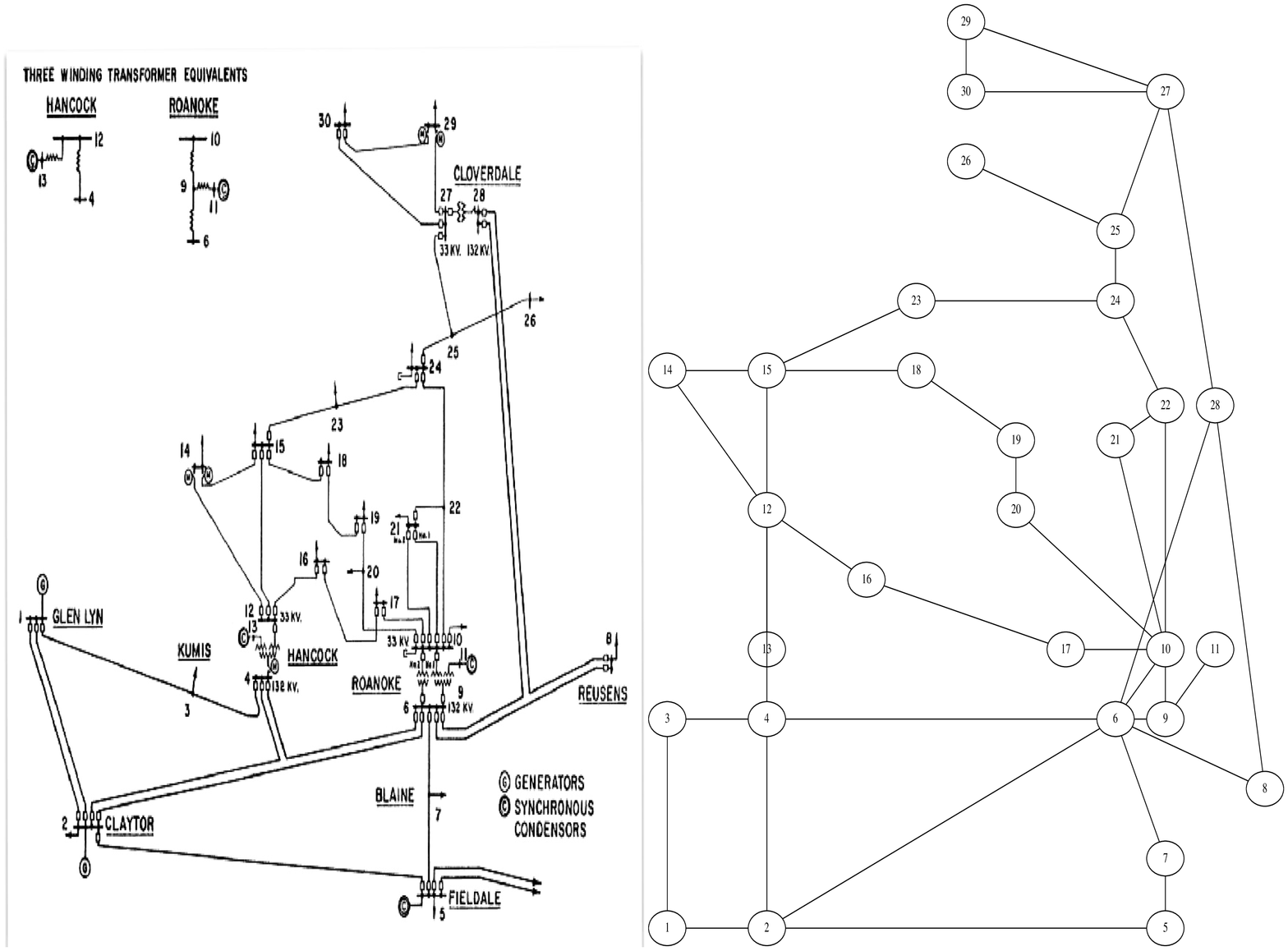,height=14cm,width=5cm,angle=90}}
\caption{Left : Electrical representation of the IEEE network case 30.
Right : schematic of the IEEE network case 30, from \cite{case30} using
the Graphviz software \cite{graphviz}.}
\label{fig_case30}
\end{figure}

The case30 network from IEEE \cite{case30} is shown in
the left panel of Fig. \ref{fig_case30}.
The graph is presented in the right panel of 
Fig. \ref{fig_case30}; it has $n=30$ vertices, 
$m=41$ edges and an average degree $\bar d = 2m/n \approx 2.7$ . 

\begin{figure}[H]
\centerline{ \epsfig{file=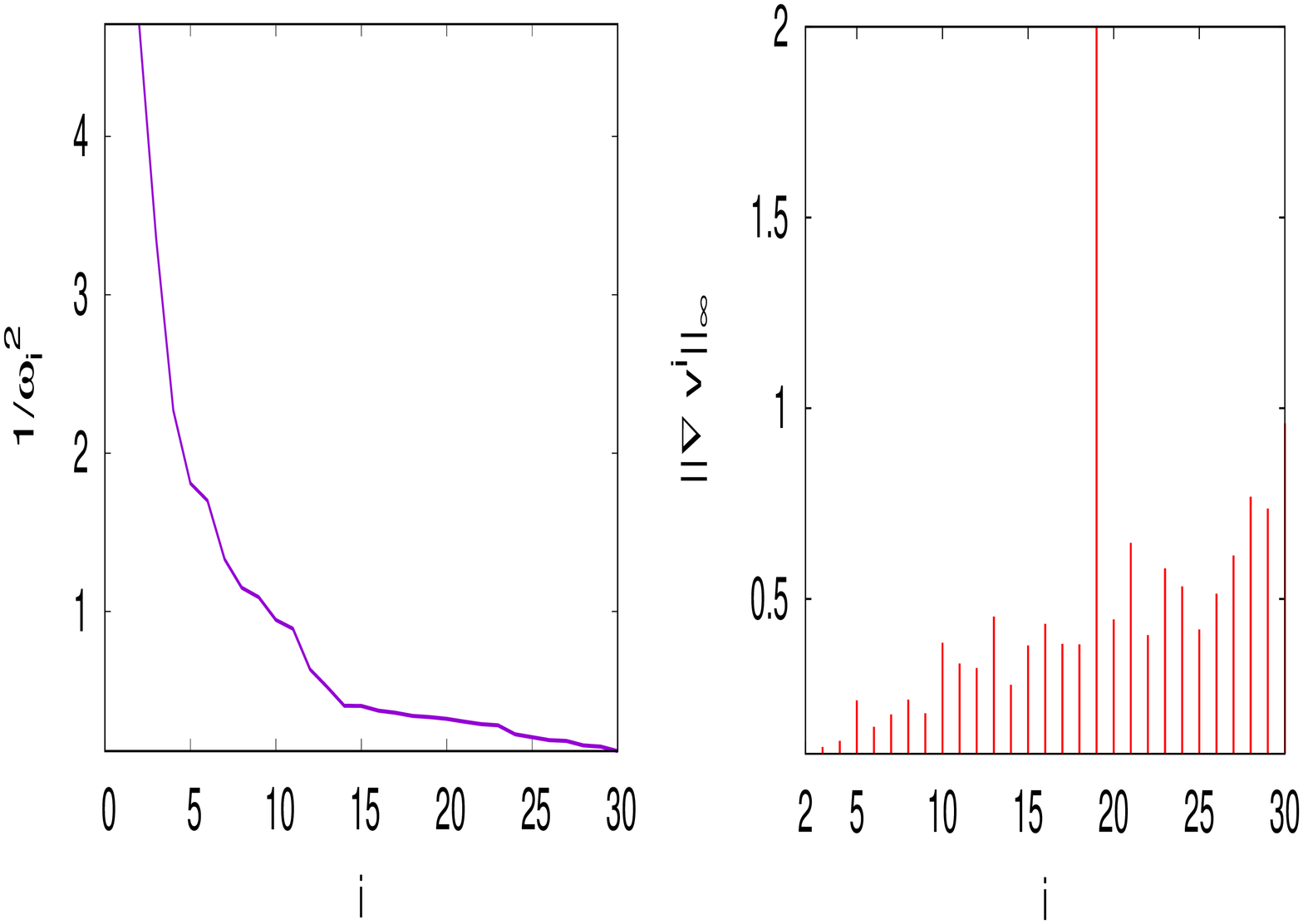,height=14cm,width=5cm,angle=90}}
\caption{
Plot as a function of $i$
of the inverse of the eigenvalue $1/ \omega_i^2$ 
(left panel) and of $ \parallel \nabla \mathbf{v}^i \parallel _{\infty}$ (right panel) .}
\label{30odnabla}
\end{figure}

For each index $i$, we compute the
inverse of the eigenvalue $1/ \omega_i^2$; it decays as a 
function of $i$ as shown in the left panel of Fig. \ref{30odnabla}.  
The norm of $ \parallel \nabla \mathbf{v}^i \parallel _{\infty}$ 
increases with $i$ and has some maxima. It is shown in the
right panel of Fig. \ref{30odnabla}.
Note the peak for $i=19$ which corresponds 
to the eigenvector $\mathbf{v}^{19}$ such that 
$\mathbf{v}^{19}_{29}=+1/\sqrt{2},~~\mathbf{v}^{19}_{30}=-1/\sqrt{2}, ~~\mathbf{v}^{19}_i=0$ 
for $i$ different from $29, 30$. The strict nodal domains
are very small, $\{29\} \cup \{30\}$.
This very special eigenvector was analyzed in
our previous work \cite{cks13}, we termed it a closed swivel because only two
nodes are non zero. On almost all nodes, no action is effective on the
system on that particular eigenmode. The eigenvalue is $\omega^2_{19}=3$.
\vskip -.3cm

\begin{figure}[H]
\centerline{\epsfig{file=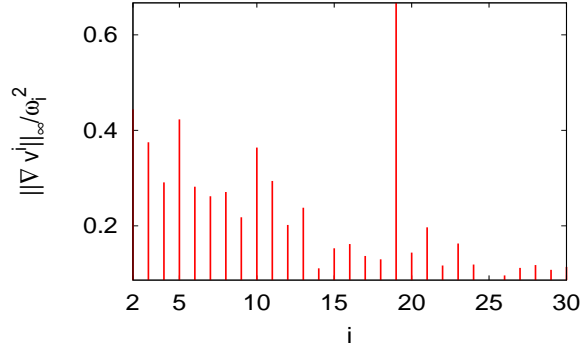,height=5cm,width=8cm,angle=0}}
\caption{
Plot of the line power infinite norm $ \parallel P_l \parallel _{\infty}$ when 
$P=\mathbf{v}^i$ as a function of $i$.
}
\label{30npl}
\end{figure}

The associated line power infinite norm 
$\parallel P_l \parallel _{\infty}$ which is the
multiplication of the two different expressions
is shown in Fig. \ref{30npl}. 
This quantity is maximum for $i=19$, corresponding 
exactly to the swivel eigenvector discussed above.
This eigenvector corresponds to the power being focused in 
the line between the two nodes of the swivel, giving the 
maximum $ \parallel P_l \parallel _{\infty}$.

\begin{figure}[H]
\centerline{
\epsfig{file=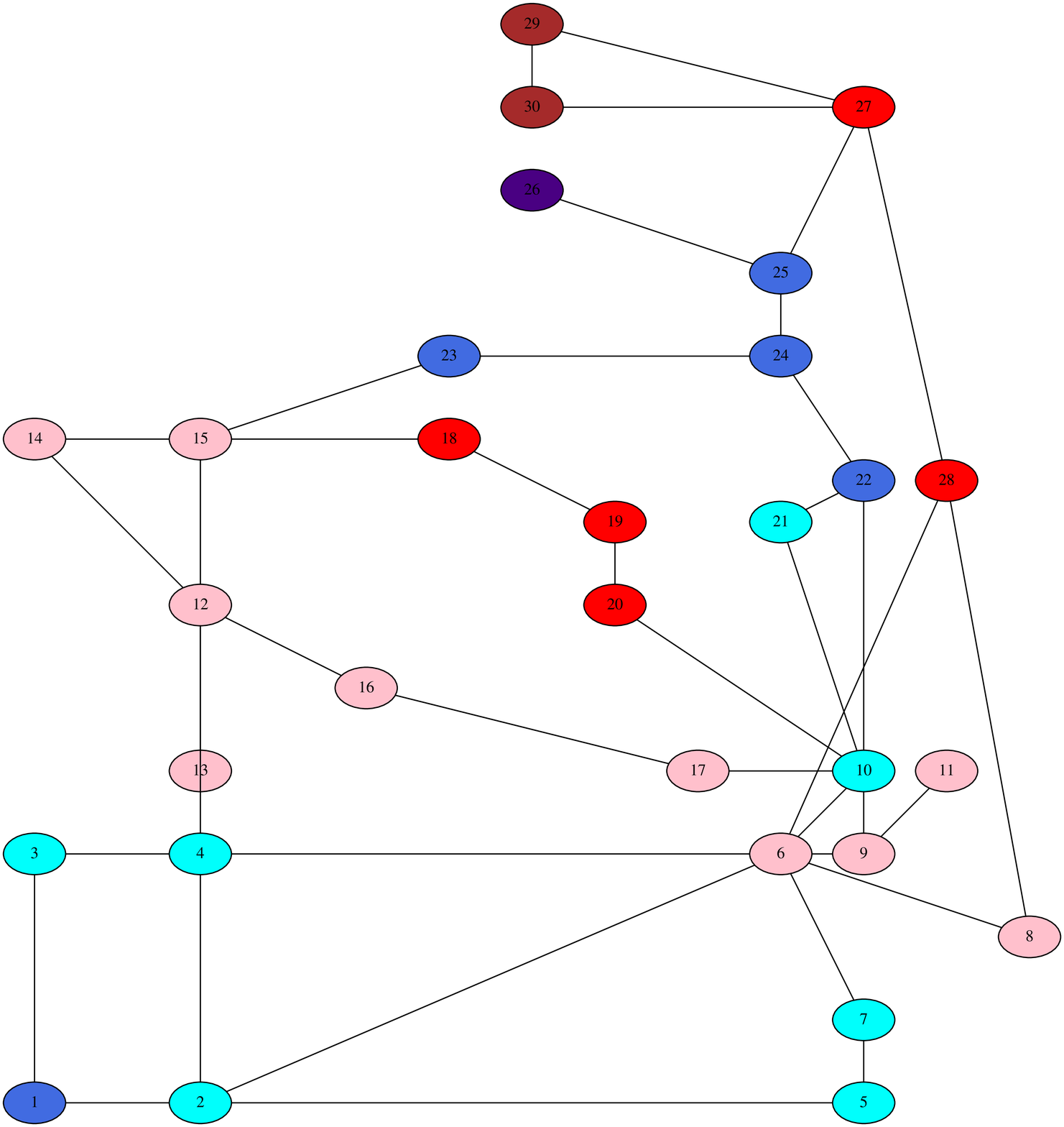,height=6cm,width=12cm,angle=0}
}
\caption{
Nodal domains of the eigenvector $\mathbf{v}^5$. 
The color scheme for the components $\mathbf{v}^5_i$ is brown if
$-0.3 < \mathbf{v}^5_i $, red if $-0.3 < \mathbf{v}^5_i < -0.1$, 
pink if $-0.1 < \mathbf{v}^5_i < 0$,
cyan if $0 < \mathbf{v}^5_i < 0.1$ , royalblue if $0.1 < \mathbf{v}^5_i < 0.3$
and indigo if $0.3 < \mathbf{v}^5_i $ . 
}
\label{v5_30}
\end{figure}

\begin{figure}[H]
\centerline{
\epsfig{file=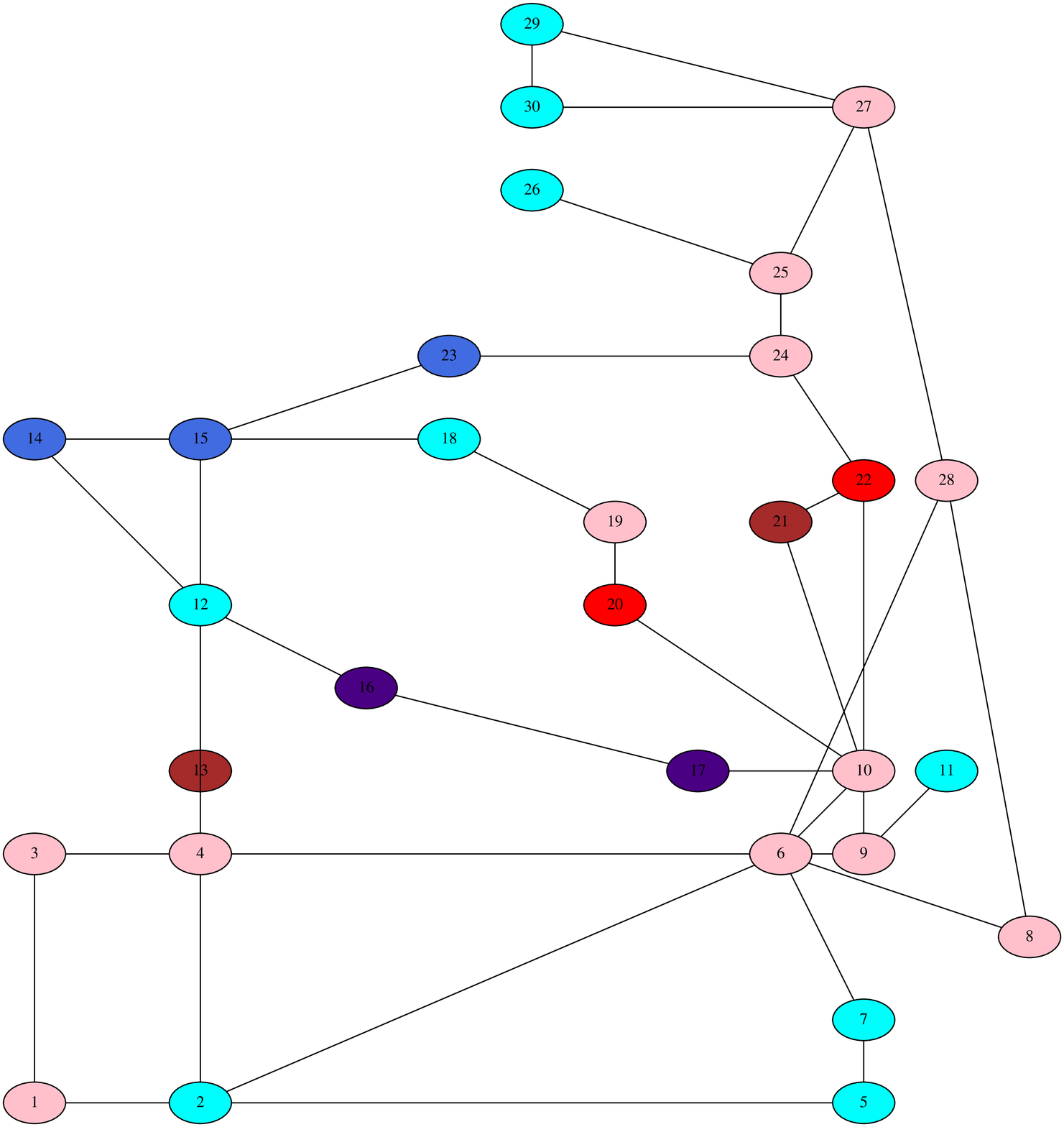,height=6cm,width=12cm,angle=0}
}
\caption{
Nodal domains of the eigenvector $\mathbf{v}^{10}$, the color scheme
is the same as for Fig. \ref{v5_30} .
}
\label{v10_30}
\end{figure}

The other eigenvectors that give peaks in 
$ \parallel P_l \parallel _{\infty}$ are
$\mathbf{v}^5$ and $\mathbf{v}^{10}$. Their nodal domains are more complex
than the one of $\mathbf{v}^{19}$ and are shown in Figs. \ref{v5_30} and \ref{v10_30}.
They both show strong gradients between nodal domains which
explain the peaks in $\parallel P_l \parallel$. From Fig. \ref{30odnabla} 
we expect that the vector $\mathbf{v}^{15}$ will contribute to $ \parallel P_l \parallel _{\infty}$, however $\omega_5^2$ is large so that finally the
contribution of $\mathbf{v}^5$ to $ \parallel P_l \parallel _{\infty}$ is small.
The positive nodal domains are 
$A=\{10,21,22,23,24,25,26\}$
$B=\{1,2,3,4,5,7\}$.
The negative nodal domains are
$C=\{6,8,9,11,28,27,29,30\}$
and  \\
$D=\{13,12,14,15,16,17,18,19,20\}$.
Note the strong gradients at the interface between the
positive and negative nodal domains. In particular
between the nodes 27 and 25 because
$-0.3 < \mathbf{v}^5_{27} < -0.1$ and 
$0.1 < \mathbf{v}^5_{25} < 0.3$.
This gradient is
responsible for the peak observed for $i=5$ in Fig. \ref{30npl}.
Notice also the strong gradient between nodes 15 and 23.

The negative nodal domains are
$A=\{1,3,4,13,6,8,9,10,19,20,21,22,24,25,27,28\}$.
The positive nodal domains are
$B=\{2,5,7\}$, $C=\{11\}$, $D=\{12,14,15,16,17,18,23\}$, 
$E=\{26\}$ and $F=\{29,30\}$.
Notice the strong gradients between nodes 10 and 17 and 10
and 21. This explains the large amplitude in 
$ \parallel \nabla \mathbf{v}^i \parallel _{\infty}$.

\begin{figure}[H]
\centerline{
\epsfig{file=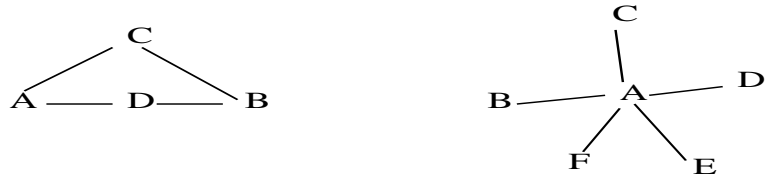,height=3cm,width=12cm,angle=0}
}
\caption{
Schematic nodal domains of the eigenvectors $\mathbf{v}^5$ (left) and  
	$\mathbf{v}^{10}$ (right).
}
\label{nodal}
\end{figure}
The comparison between $\mathbf{v}^5$ and $\mathbf{v}^{10}$ is instructive. Fig. \ref{nodal}
shows the nodal domains for $\mathbf{v}^5$ (left) and $\mathbf{v}^{10}$ (right). There are
four nodal domains for the former forming a cycle and six for the latter
forming a star. There is no general theory predicting the shape and size
of these domains, only an upper bound on their number depending on the
order of the eigenvalue.

\subsection{IEEE Case 118}

\begin{figure}[H]
\centerline{ \epsfig{file=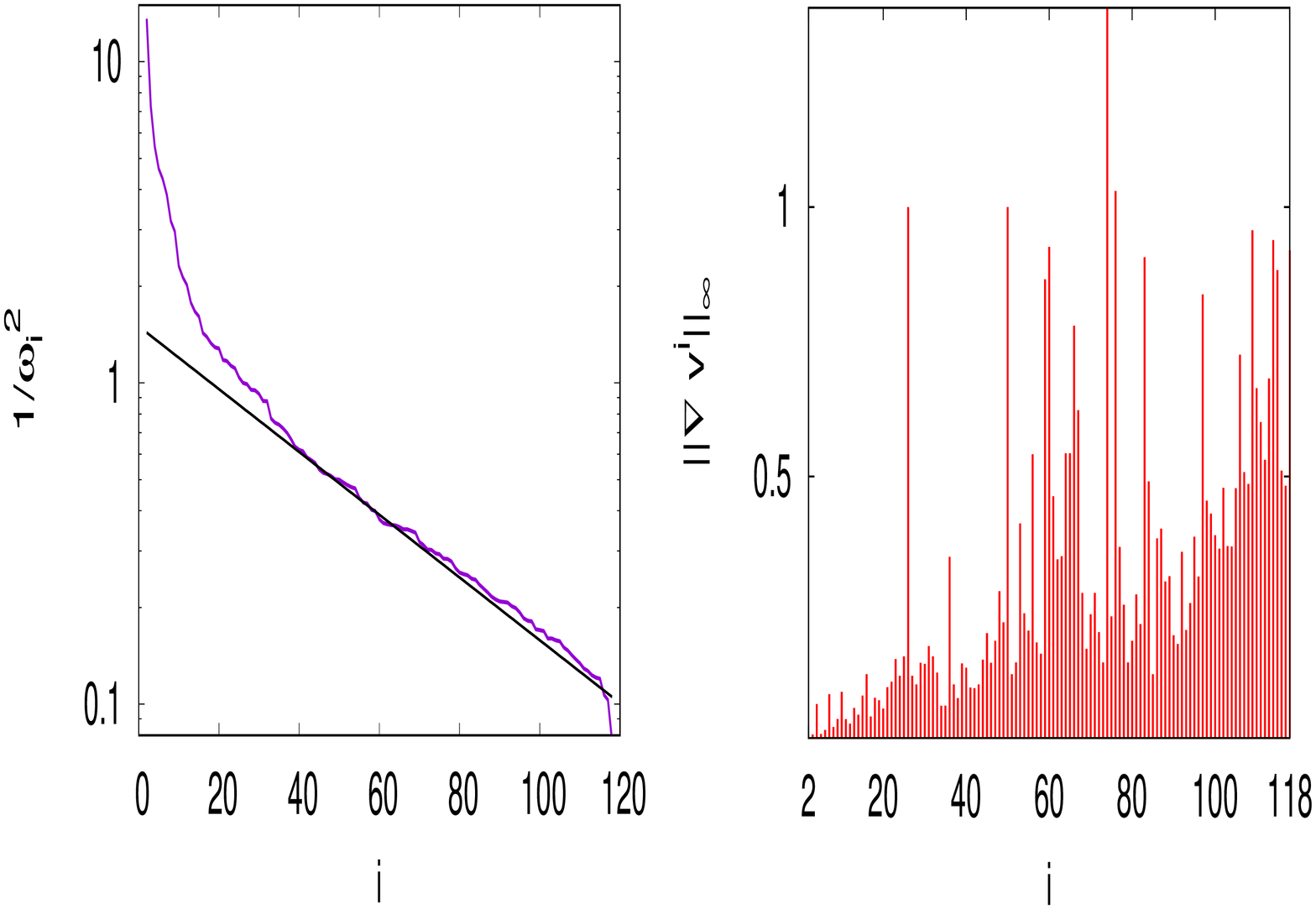,height=14cm,width=5cm,angle=90}}
\caption{
Plot as a function of $i$
of the inverse of the eigenvalue $1/ \omega_i^2$
(left panel) and of $ \parallel \nabla \mathbf{v}^i \parallel _{\infty}$ (right panel) .}
\label{118odnabla}
\end{figure}

\begin{figure}[H]
\centerline{\epsfig{file=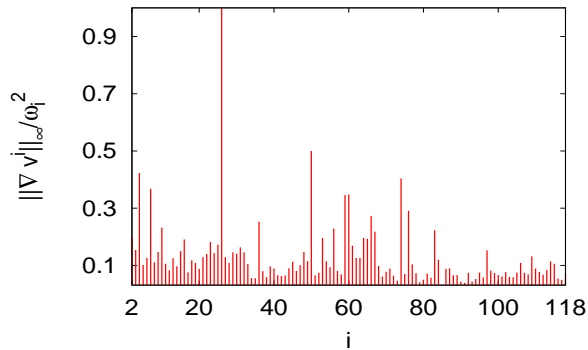,height=5cm,width=8cm,angle=0}}
\caption{
Plot of the line power infinite norm $ \parallel P_l \parallel _{\infty}$ when
$P=\mathbf{v}^i$ as a function of $i$.
}
\label{118npl}
\end{figure}

The next example is the larger case118 with $n=118$ nodes, $m=186$
edges and an average degree $\bar d = 2m/n= 3.1$. 
Note that for this network,  five lines have been doubled so that the
Laplacian now has weights. 
The evolutions of $1 / \omega_i^2$ and $\parallel \nabla \mathbf{v}^i\parallel_{\inf}$
are shown in
Figs. \ref{118odnabla} and \ref{118npl}. They are very similar
to the ones for the case30. In particular, the inverse of the
eigenvalues decay exponentially as shown in the lin-log scale of the
left panel of Fig. \ref{118odnabla}.
Notice in the right panel of Fig. \ref{118odnabla} the strong 
contributions to $ \parallel \nabla \mathbf{v}^i \parallel _{\infty}$
of the eigenvectors $\mathbf{v}^{26}, \mathbf{v}^{50}, \mathbf{v}^{59}, \mathbf{v}^{60}, \mathbf{v}^{74}, \mathbf{v}^{76}$ and 
$\mathbf{v}^{83}$.
An extreme case is the swivel eigenvector \cite{cks13}
$\mathbf{v}^{26}$ such that $\mathbf{v}^{26}_{111}=+1/\sqrt{2},~~\mathbf{v}^{26}_{112}=-1/\sqrt{2}, ~~\mathbf{v}^{26}_i=0$ for $i$ 
different from $111,~112$. The eigenvalue is $\omega^2_{26}=1$.
The eigenvector $\mathbf{v}^{50}$ is also a swivel. 
The other eigenvectors are localized in specific regions of the network.
By this we mean that the eigenvector has a small number of components of
absolute value much larger than the rest. For example
$\mathbf{v}^{59}$ is localized from nodes 84 to 88. $\mathbf{v}^{60}, \mathbf{v}^{83}$ from 100 to 118,
$\mathbf{v}^{74}$ around 90 and $\mathbf{v}^{76}$ around 110.
This localization comes as a surprise because 
the general theory of nodal domains does not predict it.

The associated line power infinite norm $ \parallel P_l \parallel _{\infty}$ 
is shown in Fig. \ref{118npl}. Not all the peaks present in the
right panel of Fig. \ref{118odnabla} are present here. This is because
of the increase of the eigenvalues $\omega^2_i$ with $i$. For example
the large peaks $\mathbf{v}^{74},\mathbf{v}^{76}$ are now much smaller in Fig. \ref{118npl}.
The swivel eigenvector $\mathbf{v}^{26}$ gives the largest contribution.

To conclude this section, we have seen that 
$\nabla \mathbf{v}^i$ is related to nodal domains. We see a 
general trend showing that a linear interpolation of 
$\parallel \nabla \mathbf{v}^i \parallel$ shows a slow increase with $i$. 
However there are 
some peaks  that correspond to highly localized eigenvectors.
These highly localized eigenvectors $\mathbf{v}^i$ give a large contribution
to $\nabla \mathbf{v}^i$. Some are due to geometrical configurations of the
network like swivels. It is not clear where the others arise from.

In the next section, we consider general $P$ distributions. We will
see that localized eigenvectors play an important role in $P_l$ for small $i$.
When $i$ is large, their influence is mitigated by the denominator
$\omega_i^2$.

\section{Spectral solutions of the reduced load-flow}

In this section, we combine the graph information with
the generator / load vector and calculate the
power in the lines $P_l$.

\subsection{A small size network : effect of soft nodes} 

Before addressing networks with a relatively large number of nodes
it is useful to consider a very simple example where calculations
can be conducted by hand. This shows the usefulness of the approach.

\begin{figure}[H]
\centerline{
\epsfig{file=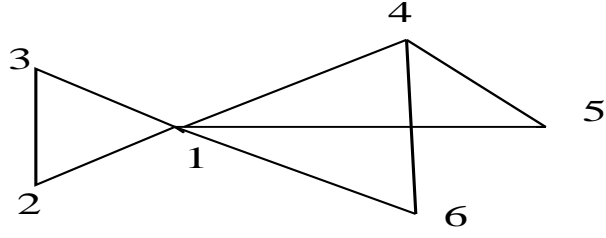,height=3cm,width=8cm,angle=0}
}
\caption{
A 6-node electrical network.}
\label{61}
\end{figure}

We consider the simple 6 node network shown in Fig. \ref{61}.

The graph Laplacian here is
\be\label{lap61} \Delta = \begin{pmatrix}
   5 &  -1 &  -1  & -1  & -1 & -1 \cr
  -1  &  2  & -1  & 0  &  0  &  0 \cr
  -1  & -1  & 2   & 0  & 0   &  0 \cr
  -1  & 0   &  0  & 3  & -1  & -1 \cr
  -1  & 0   &  0  & -1 &  2  & 0 \cr
  -1  & 0   & 0   & -1 &  0  &  2 
\end{pmatrix}, \ee
whose eigenvalues $\omega_i^2,~i=1,\dots,6$ are
\be\label{eigval61}
0 , ~~~1,~~~2,~~~3, ~~~4,~~~6 \ee
corresponding to the eigenvectors 
\begin{eqnarray}\label{vlap61} 
\mathbf{v}^1 = {1 \over {\sqrt{6}}} (1,1,1,1,1,1)^T, 
~~~~\mathbf{v}^2={1 \over \sqrt{30}} (0,3,3,-2,-2,-2)^T,\nonumber \\
\mathbf{v}^3={1 \over \sqrt{2}} (0,0,0,0,1,-1)^T,
~~~~\mathbf{v}^4 = {1 \over {\sqrt{2}}} (0,1,-1,0,0,0)^T, \nonumber \\
\mathbf{v}^5={1 \over \sqrt{6}} (0,0,0,2,-1,-1)^T,
~~~~\mathbf{v}^6={1 \over \sqrt{30}} (-5,1,1,1,1,1)^T,\nonumber 
.  \end{eqnarray}
The associated gradients are 
\begin{table} [H]
\centering
\begin{tabular}{|l|r|}
\hline
$\nabla \mathbf{v}^2$  & $(-0.55,0,0.55,0.36,0,-0.36,0.36,0)^T$ \\
$\nabla \mathbf{v}^3$  & $(0,0,0,0,-0.71,0.71,0.71,0.71)^T$, \\
$\nabla \mathbf{v}^4$  & $(0.71,-1.41,0.71,0,0,0,0,0)^T$, \\
$\nabla \mathbf{v}^5$  & $(0,0,0,0.82,-1.22,0.41,-0.41,-1.22)^T$, \\
$\nabla \mathbf{v}^6$  & $(-1.09,0,1.09,-1.09,0,1.09,-1.09,0)^T$  \\ \hline
\end{tabular}
\label{tab2a}
\end{table}

The power in the lines is then 
\be\label{pl61}
P_l = p_2 { \nabla \mathbf{v}^2 \over 1}
+ p_3 { \nabla \mathbf{v}^3 \over 2} 
+ p_4 { \nabla \mathbf{v}^4 \over 3}
+ p_5 { \nabla \mathbf{v}^5 \over 4}
+ p_6 { \nabla \mathbf{v}^6 \over 6} ,\ee
where $p_i$ is the projection of $P$ on the eigenvector $\mathbf{v}^i$, see
(\ref{proj_p}).
Expression (\ref{pl61}) suggests that a large $p_2$ will contribute 
significantly more to $P_l$ than a large $p_5$ or $p_6$.

When the eigenvector $\mathbf{v}^i$ has a zero component at node $k$, 
$\mathbf{v}^i_k=0$ ( a soft node in the language of \cite{cks13}), the $p_i$ 
coefficient does not depend on what is at node $k$. This is because
$p_i= P \cdot \mathbf{v}^i$. In particular, if there is a generator at node $k$, 
it will not contribute to $p_i$. This reduces the number of directions
for minimizing $\parallel P_l \parallel$.

To see these effects in more detail, we first assume that the loads 
are equally distributed over the network
and study how placing a single generator on the network affects $P_l$.
To examine the contribution of the different modes $\mathbf{v}^i$ 
to $P_l$, we introduce the partial sums
\be\label{part_pli}
s_k^\infty =  \parallel \sum_{i=2}^k (g_i-l_i) {\nabla \mathbf{v}^i \over \omega_i^2} \parallel _\infty,\ee
\be\label{part_pl2}
s_k^2 = \sum_{i=2}^k {(g_i-l_i)^2 \over \omega_i^2}.\ee
Note that $s_n^2=  \parallel P_l \parallel ^2_2$ and $s_n^\infty= \parallel P_l \parallel _\infty$. 
\begin{table} [H]
\centering
\begin{tabular}{|c|c|c|c|c|}
   \hline
position      &      &      &      & \\
of generator  & 1    &  2   &  4   &  6 \\ \hline
$p_2$         & 0    &-3.29 & 2.19 &  2.19  \\ \hline
$p_3$         & 0    &  0   &  0   &-4.24   \\ \hline
$p_4$         & 0    & 4.24 &  0   & 0    \\ \hline
$p_5$         & 0    & 0    &  5   & -2.44  \\ \hline
$p_6$         & 5.48 &-1.09 & -1.09&  -1.09  \\ \hline
$ \parallel P_l \parallel _\infty$ & 1 & 3    & 2    &  2.75  \\ \hline
$ \parallel P_l \parallel _2$      & 2.24 & 4.12   & 3.32  &  3.94  \\ \hline
\end{tabular}
\caption{Power coefficients $p_i$, $ \parallel P_l \parallel _\infty$, 
$ \parallel P_l \parallel _2$
for different generator positions. The loads are uniformly distributed.
}
\label{tab3}
\end{table}
Table \ref{tab3} shows the coefficients $p_i$ for a generator
of strength 6 placed at nodes 1, 2 or 4. A generator at node 1
will be such that only $p_6$ is non zero. Then we expect that
$ \parallel P_l \parallel $ will be minimal and this is indeed the case.
On the other hand, a generator placed at node 2 gives
a large $p_2$ so that $ \parallel P_l \parallel $ will be larger. 
As expected, we see in table \ref{tab3} 
a correlation between large values of $p_2$ and $p_3$
and large values of $ \parallel P_l \parallel $.

\begin{figure}[H]
\centerline{ \epsfig{file=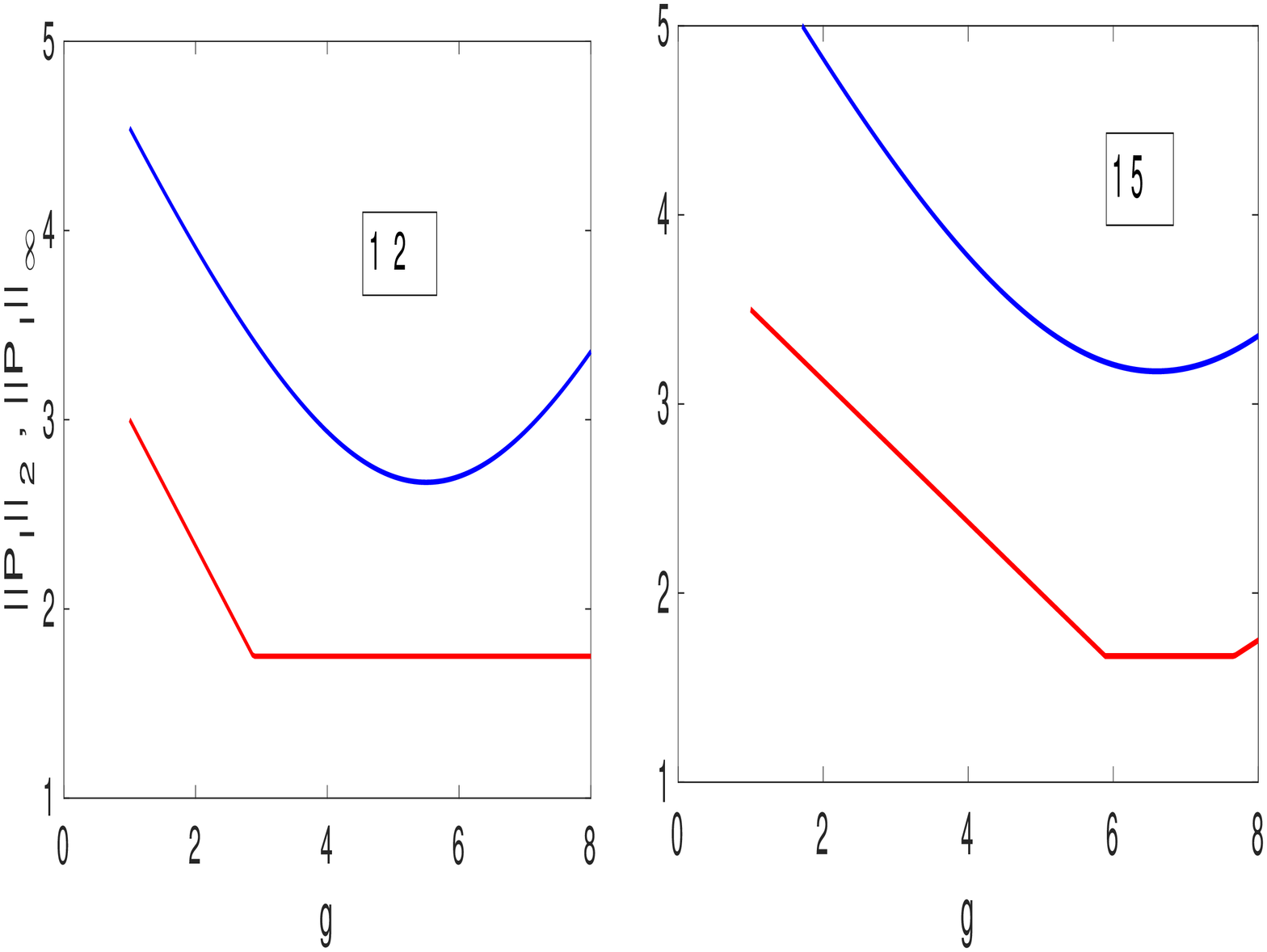,height=14cm,width=5cm,angle=90}}
	\caption{ Plot of $ \parallel P_l \parallel _2$ (blue online) and $ \parallel P_l \parallel _\infty$
	(red online) as a function of the strength $g$ of the generator at
	node 1, when the second generator is placed at node 2
	(left panel) or at node 5 (right panel).}
\label{pl61a}
\end{figure}

In a second set of experiments, we place two generators on the
grid and examine how $P_l$ depends on their position. For this, 
we choose the following vector of loads
$$L = (1,2,1,3,0,1)^T . $$
First we assume that the generators are placed at nodes 1 and 2, so that
$G = ( G_1, G_2,0,0,0,0)^T$, where $G_1+G_2= \sum_i L_i$. Then the
power vector is $P = ( g-1, 6-g, -1, -3, 0, -1)^T$ where we replaced
$G_1$ by $g$ to simplify the notation. In the
following, Projecting
$P$ onto the eigenvectors, we note that, because of the zero components
$\mathbf{v}^3_1$ and $\mathbf{v}^5_1$, 
there are no $g$ dependent
components on the eigenvectors $\mathbf{v}^3$ and $\mathbf{v}^5$; we find
$ \parallel P_l \parallel _2 = 2.7$. When the generators are now placed at nodes
1 and 5, $g$ terms will affect the components of $\mathbf{v}^2, \mathbf{v}^3, \mathbf{v}^5$
and $\mathbf{v}^6$. We then expect to find a higher maximum for $ \parallel P_l \parallel _2$
and this is the case, $ \parallel P_l \parallel _2 = 3.2$. Fig. \ref{pl61a}
shows $ \parallel P_l \parallel _2$ (blue online) and $ \parallel P_l \parallel _\infty$ (red online)
as a function of $g$ for the two different configurations. 
We see that $ \parallel P_l \parallel _2$ for the 1-2 configuration (left) is
always above $ \parallel P_l \parallel _2$ for the 1-5 configuration (right). On
the other hand, the minimum of $ \parallel P_l \parallel _\infty$ is the same for
both configurations. 
The flatness of $ \parallel P_l \parallel _\infty$ for the 1-2 distribution
(left of Fig. \ref{pl61a}) is due to the zero first and
second components for $\mathbf{v}^i$. On the other hand the 1-5
distribution has less zeros so the $ \parallel P_l \parallel _\infty$ depends
more on $g$. Fig. \ref{pl61a} also shows that for both configurations 
$1,2$ and $2,5$, we can simultaneously minimize the two norms.

\begin{figure}[H]
\centerline{ \epsfig{file=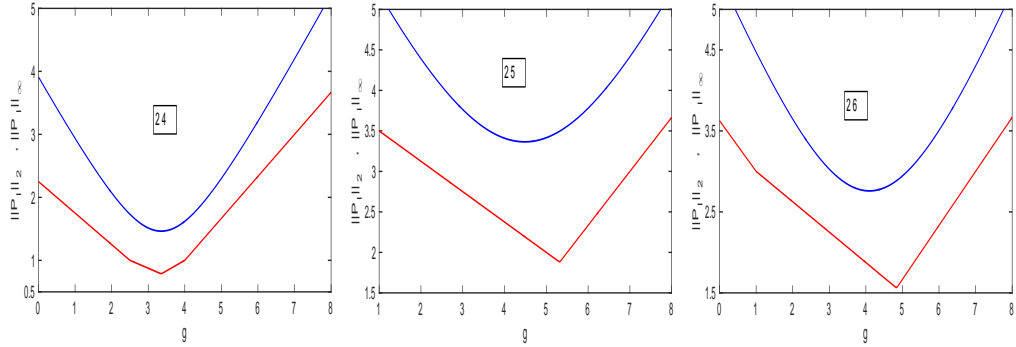,height=14cm,width=5cm,angle=90}}
        \caption{ Plot of $ \parallel P_l \parallel _2$ (blue online) and $ \parallel P_l \parallel _\infty$
        (red online) as a function of the strength $g$ of the generator at
        node 2, when the second generator is placed at nodes 4,5 and 6.}
\label{pl61b}
\end{figure}

We now place the 1st generator of amplitude $g$ at node 2
and the second one at nodes 4,5 and 6 respectively. Fig. \ref{pl61b}
shows $ \parallel P_l \parallel _2$ $ \parallel P_l \parallel _2$ (blue online) and $ \parallel P_l \parallel _\infty$
(red online) as a function of $g$.
We see that the 2-4 configuration gives a minimum compared to the 2-5 and 2-6.
This is clear because in this configuration, the $\mathbf{v}^3$ component
of $P$ does not depend on $g$. Here, only the $2,4$ configuration
(left of Fig. \ref{pl61b})
leads to the same minimum for $ \parallel P_l \parallel _2$
and  $ \parallel P_l \parallel _\infty$.

\subsection{Convergence of $s_{k}^2$ with $k$: example of a grid}

For the placement of two generators on a network, it is interesting
to write $\parallel P_l \parallel _2^2$. Assuming the generators are
positioned at nodes $p$ and $m$ , with amplitudes $G_p$ and $G_m$, we have
$$\parallel P_l \parallel _2^2 = \sum_{i=2}^n {(g_i -l_i)^2 \over \omega_i^2} =
\sum_{i=2}^n {(G_p v_p^i + G_m \mathbf{v}^i_m -l_i)^2 \over \omega_i^2}.$$
Expanding the squares and rearranging, we get the final expression
$$\parallel P_l \parallel _2^2 = G_p^2 \sum_i {(v_p^i)^2 \over \omega_i^2}
+ G_m^2 \sum_i {(v_m^i)^2 \over \omega_i^2}$$
\be\label{pl2km}
+ \sum_i {l_i^2 \over \omega_i^2}
+ 2 G_p G_m \sum_i {v_p^i v_m^i \over \omega_i^2}
- 2 G_p \sum_i {v_p^i l_i \over \omega_i^2}
- 2 G_m \sum_i {v_m^i l_i \over \omega_i^2} . \ee
The coefficients of this polynomial in $G_p, G_m$ are sums from $i=2$
to $n$. We have observed that they converge rapidly with $i$.

Simple systems on which to test this convergence are chains and
grids (cartesian product of two chains). There, one can compute
explicitly the eigenvectors and eigenvalues so that the
network can be made arbitrarily large. A grid is also 
a first approximation of a transmission network.

A chain with $n$ nodes has eigenvalues $\omega^2_{i}$ and 
eigenvectors $\mathbf{v}^i$ whose components $\mathbf{v}^{i}_{p}$ are
\be\label{gli}
\omega^2_{i} = 4 \sin^2 {\pi (i-1) \over 2 n} , ~~i=1,\dots, n\ee
\be\label{vip}
\mathbf{v}^{i}_{p}= 
{1 \over N_i }
\cos [{\pi (i-1) \over  n}(p -{1\over 2}) ], ~~p=1,\dots, n \ee
where the normalization factor is $N_i = \sqrt{n}$ if $i=1$
and $N_i = \sqrt{n/2}$ otherwise.

Let us consider $ \parallel P_l \parallel^2_2$ for this network. From 
\ref{pl2} we have
$$ \parallel P_l \parallel^2_2 = \sum_{i=2}^n {p_i^2 \over \omega_i^2}=
{1\over 4} \sum_{i=2}^n {p_i^2 \over \sin^2 {\pi (i-1) \over 2 n}}$$
The error committed when truncating the sum at $k \le n $ is
$$\delta_{k} \equiv \parallel P_l \parallel^2_2 - s_{k}^2
= {1\over 4} \sum_{i=k}^{n-1} {p_{i+1}^2 \over \sin^2 {\pi i \over 2 n}} .$$
This quantity is positive and the sequence 
${1 \over \sin^2 {\pi i \over 2 n}}$ is decreasing so that
$\delta_{k}$ has the following upper bound
$$\delta_{k} \le {1\over 4}~~ \underset {k+1 \le i\le n} {\rm max}~ (p_i^2)~~
\int_{k-1}^{n-1} {dx \over \sin^2 {\pi x \over 2 n}}.$$
Finally we obtain
\be\label{bsk2} \delta_{k} \le {n \over 2 \pi}~~  
\underset {k+1 \le i\le n} {\rm max}~
( p_i^2) ~~
\left[  \cotan{\pi (k-1) \over 2 n} -\cotan{\pi (n-1) \over 2 n} \right ].\ee
\begin{figure}[H]
\centerline{ \epsfig{file=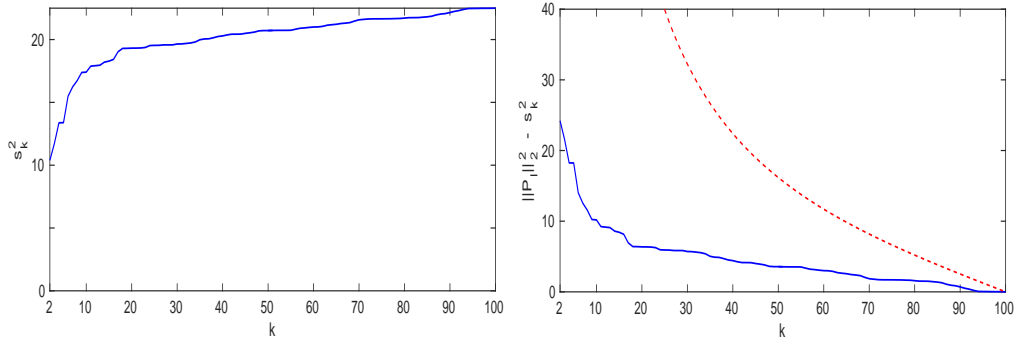,height=14cm,width=5cm,angle=90}}
\caption{
Plot of the partial sum $s_{k}^2$ (left) and the error $\delta_{k}$
(right) as a function of $k$ for a chain with $n=100$ nodes. 
The upper bound (\ref{bsk2}) is shown in dashed
line (red online). See text for parameters.}
\label{ch10}
\end{figure}
To see how good the estimate (\ref{bsk2}) we studied a chain
with $n=100$ nodes. The generator vector is such that
$G(31)=1,~G(5)=3$, $0$ elsewhere
and the load vector verifies $L(4)=2,~ L(62)=1,~L(15)=1$ and $0$ elsewhere.
The left panel of Fig. \ref{ch10} shows the partial sum $s_{k}^2$ as
a function of $k$. It reaches 80 \% of its value for $k\approx n/5$.
The error $\delta_{k}$ (right panel) decreases sharply for $k < n/5$,
afterwards its decrease is much slower. The upper bound (\ref{bsk2}) 
is shown in dashed line (red online). The fairly large difference is
due to $p_i^2$. This quantity depends on the eigenvectors and is difficult
to estimate; the only option is to take the upper bound 
$ \underset {k+1 \le i\le n} {\rm max} p_i^2 $. We will discuss this
at the end of the section.

Consider now a grid formed by the cartesian product 
$C_n \times C_m$ of two chains $C_n$ and $C_m$ with $n$ and $m$ 
nodes respectively. Its eigenvalues are
$\omega^2_{i,j} = \omega^2_i + \omega^2_j $ where $\omega^2_i$ is an eigenvalue
for $C_n$ while $\omega^2_j$ is an eigenvalue
for $C_m$. The associated eigenvector is $\mathbf{v}^{ij} = \mathbf{v}^i \otimes \mathbf{v}^j$,
the Kronecker product of $\mathbf{v}^i$ and $\mathbf{v}^j$ (more details can be
found in the book \cite{booklapla}).
The eigenvalue $\omega^2_{i,j}$ and the components $\mathbf{v}^{ij}_{pq}$ are
\be\label{glij}
\omega^2_{i,j} = 4 \left [  
\sin^2 {\pi (i-1) \over 2 n} + \sin^2 {\pi (j-1) \over 2 m}  \right ] , \ee
\be\label{vijpq}
\mathbf{v}^{ij}_{pq}= \mathbf{v}^i_p \mathbf{v}^j_q = 
{1 \over N_p N_q}
\cos [{\pi (i-1) \over  n}(p -{1\over 2}) ]
\cos [{\pi (j-1) \over  m}(q -{1\over 2}) ], \ee
where $ i,p \in \{1,\dots ,n\},~~~j,q \in \{1,\dots ,m\}$ and where the
normalization factors $N_p,N_q$ follow the rules of the chains.
\begin{figure}[H]
\centerline{ \epsfig{file=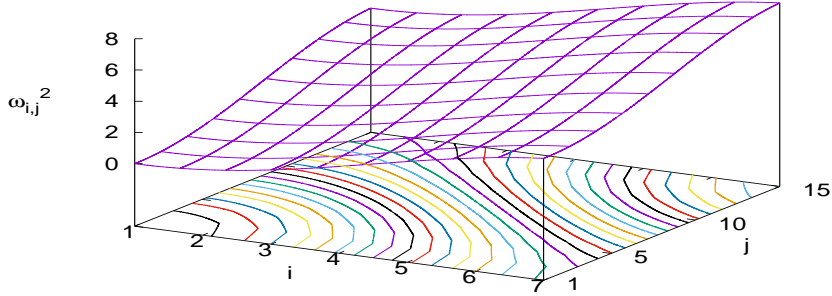,height=6cm,width=12cm,angle=0}}
\caption{ Eigenvalues $\omega^2_{i,j}$ as a function
of $i,j$ for a grid $n=7, m=15$. On the bottom we 
show the level sets ranging from $0$ to $8$ and separated by $0.25$.}
\label{lgril}
\end{figure}
The eigenvalues $\omega^2_{i,j}$ are such that 
$\omega^2_{i,j} \le 8$. They increase monotonically with $i$ and $j$ as
shown in Fig. \ref{lgril}; there the contour lines are separated by $0.25$.

\begin{figure}[H]
\centerline{ \epsfig{file=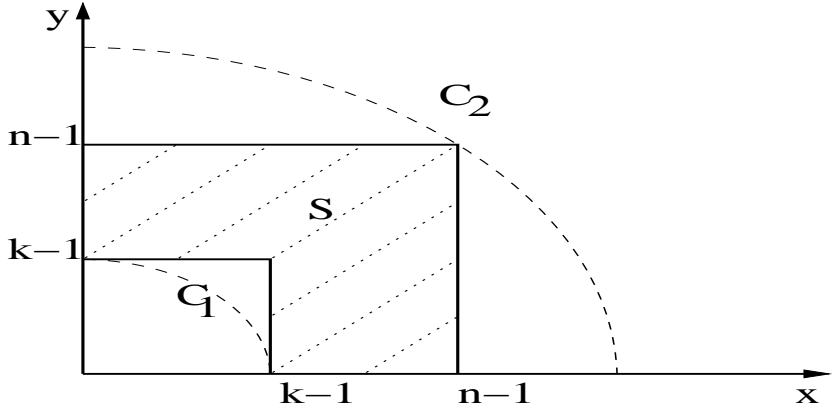,height=6cm,width=12cm,angle=0}}
\caption{ Integration domain for $I_2$ in the $(x,y)$ plane.} 
\label{anul}
\end{figure}

The expression of $\parallel P_l\parallel_2^2$ is
\be\label{pl2grid}
\parallel P_l\parallel_2^2 = \sum_i^{n} \sum_j^{n} 
{  p_{ij}^2 \over \omega^2_{ij}} , \ee
where $p_{ij}$ is the component of the power on the eigenvector
$\mathbf{v}^{ij}$ and where $p_{11}=0$. The sum is written so for ease of notation, 
the term $i=j=1$ should be omitted because $\omega_{11}=0$.
Let us consider the residual 
\be\label{res2d}
\delta_{k,l} \equiv 
\parallel P_l\parallel_2^2 - \sum_i^{k} \sum_j^{l} {  p_{ij}^2 \over \omega^2_{ij}} .
\ee
Assume for simplicity $n=m,~~k=l$. We have
$$\delta_{k,k} \le  {1\over 4} ~~
\underset {k+1 \le i,j \le n} {\rm max} (p_{ij}^2)  ~~
\sum_{i,j=k}^{n} {1 \over \sin^2 {\pi i \over 2 n}+ \sin^2 {\pi j \over 2 n} } 
\le {1\over 4} ~~
\underset {k+1 \le i,j \le n} {\rm max} (p_{ij}^2)  ~~I_2(k) ,
$$
where $I_2$ is the integral over the strip $S$, see Fig. \ref{anul}
\be\label{i2}
I_2(k) = \iint_{S}
{dx dy \over \sin^2 {\pi x \over 2 n}+ \sin^2 {\pi y \over 2 n} }
\ee
The integrand in $I_2$ is positive so $I_2$ can be bounded from above by
the integral on the quarter annulus $A$ bounded by the circles $C_1$
and $C_2$ shown in Fig. \ref{anul}.
We have $$I_2= ({2n \over \pi})^2 
\iint_{{\pi(k-1) \over 2n} \le w,z \le {\pi(n-1) \over 2n}} 
{dw dz  \over \sin^2 w + \sin^2 z } .$$
The function $ \sin^2 (r\cos \theta) + \sin^2 (r\sin \theta)$ is
minimum for $\theta=\pi/4$ so that
$${1 \over \sin^2 (r\cos \theta) + \sin^2 (r\sin \theta)}
\le {1 \over 2 \sin^2 (r/\sqrt{2})} .$$
Then
$$\delta_{k,k} \le  ~~
({n \over \pi})^2 
\underset {k+1 \le i,j \le n} {\rm max} (p_{ij}^2 )~~
\int_{0}^{\pi/2} d\theta \int_{\pi(k-1) \over 2n}^{\pi(n-1) \over n\sqrt{2}} 
{rdr \over \sin^2 {\pi r \over \sqrt{2}  }},$$
and further calculations yield the final result
\be\label{dkk}
\delta_{k,k} \le n^2 \underset {k+1 \le i,j \le n} {\rm max} (p_{ij}^2 )~~
\left[  \cotan{\pi (k-1) \over 2 \sqrt{2} n} -\cotan{\pi (n-1) \over 2 n} 
\right ].\ee
The dominant term is the first $\cotan$. It is large for $k$ small
and decays quickly as $k$ increases. For ${k-1 \over 2 \sqrt{2} n}=0.2$
$\cotan{\pi (k-1) \over 2 \sqrt{2} n} \approx  1.37$.
Again, this upper bound is not sharp because of the crude
bound on $p_{ij}^2$.

\begin{figure}[H]
\centerline{ \epsfig{file=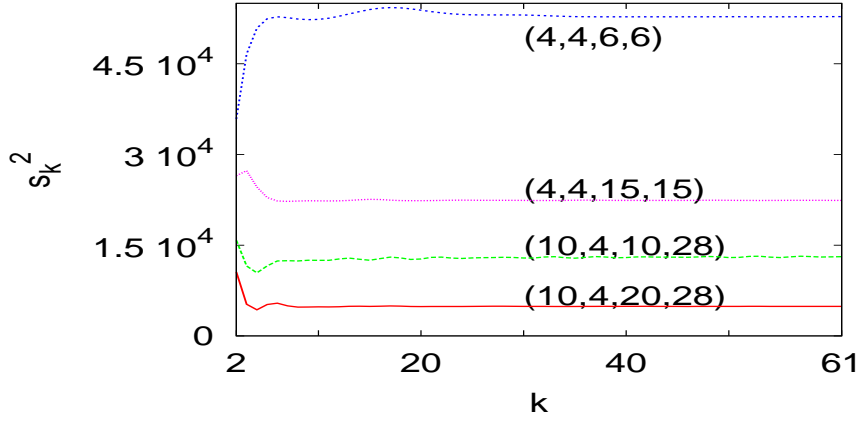,height=6cm,width=12cm,angle=0}}
\caption{ Partial sums $s_{k}^2=\sum_{ij}^{k} {\mathbf{v}^{ij}_{pq} \mathbf{v}^{ij}_{rs} \over \omega^2_{i,j}},$ as function of $k$ for different $(p,q,r,s)$
	configurations. }
\label{grins}
\end{figure}

To analyze the effects of $p_{ij}^2$, we have to fix the
distribution of generators and loads. Assume as in the
beginning of the section that we only have two generators placed
at nodes $p$ and $m$ and uniform loads. Then, we can use expression
(\ref{pl2km}) for $\parallel P_l \parallel_2^2 $. 
For the grid, the indices $i,m$ are associated to four
indices $(p,q), ~~(r,s)$. This means that we place one generator at
position $(p,q)$ and another at $(r,s)$.
Assume these positions are fixed; we introduce the partial sum
\be\label{skg2}
s_{k}^2=\sum_{i,j =1}^{k} {\mathbf{v}^{ij}_{pq} \mathbf{v}^{ij}_{rs} \over \omega^2_{i,j}},\ee
with the restriction that we omit the term $i=j=1$.
To examine how $s_{k}^2 \to s_{n}^2$, 
we considered a grid of size $n=61, m=61$ and computed $s_{k}^2$
for $(p,q,r,s)=(10,4,20,28), (10,4,10,28), 
(4,4,6,6)$ and $(4,4,15,15)$. The results are shown in Fig. \ref{grins}.
In all cases, except for the close nodes configuration
$(4,4,6,6)$, the sum converges for $k \approx 10$. For the $(4,4,6,6)$
the sum has converged for $k \approx 20 \ll n$.
We observe similar fast convergence of the other sums
in expression (\ref{pl2km}).

\subsection{The IEEE 30 network} 

There are only six generators in this network,
\be \label{gen_case30}
G_1 = 23.54, ~G_2= 60.97, ~G_{13}=37, ~ G_{22}=21.59,~ G_{23}=19.2,~ G_{27}=26.91.
\ee
The loads are distributed uniformly over the network.
\begin{figure}[H]
\centerline{ \epsfig{file=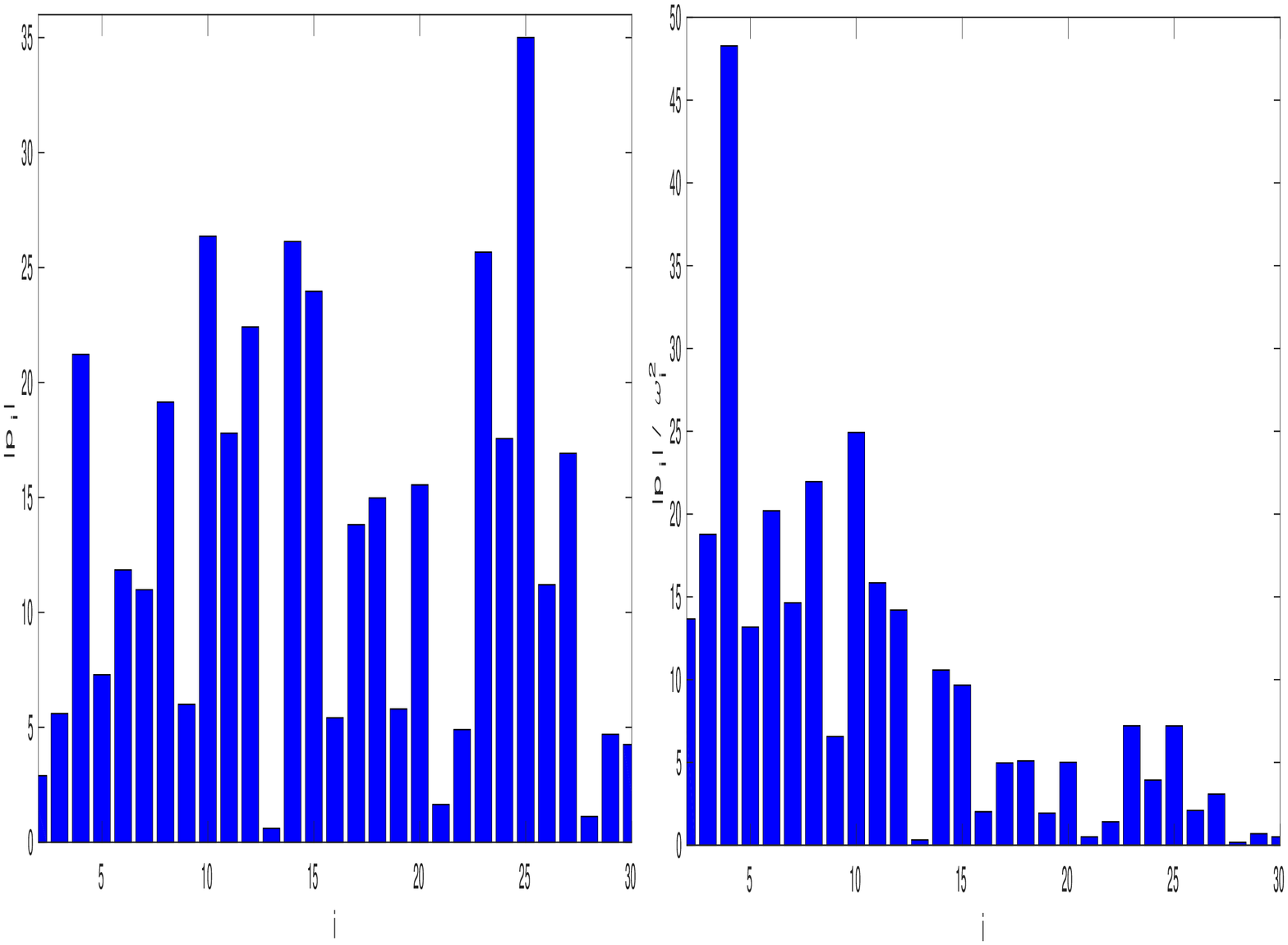,height=14cm,width=5cm,angle=90}}
\caption{
Plot of $|p_i|$ (left) and ${|p_i| \over \omega_i^2}$ (right) as a function of $i$ for the power vector $P$ of
IEEE case 30.}
\label{pp30}
\end{figure}
The components of the power vector $P$ are shown in Fig. \ref{pp30}. As
shown in the right panel, 
The right panel shows that, as expected, ${|p_i| \over \omega_i^2}$ 
decays with $i$.

\begin{figure}[H]
\centerline{\epsfig{file=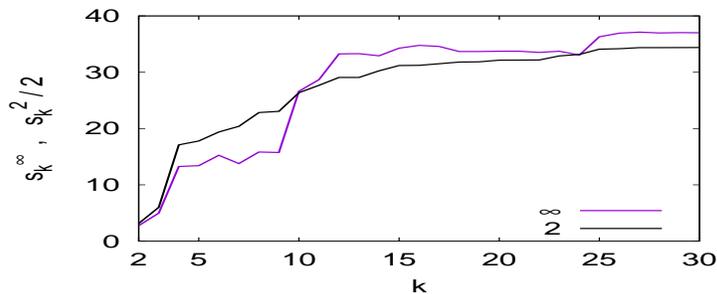,height=4cm,width=10cm,angle=0}}
\caption{
	Plot of the partial sums $s_{k}^\infty$, $s_{k}^2/2$ from 
	(\ref{part_pli},\ref{part_pl2}) as a function of ${k}$.}
\label{pld2_30}
\end{figure}

First, we examine the convergence of $s_k^\infty, ~~s_k^2$ as $k$
increases. The graph is shown in Fig. \ref{pld2_30}.
Note how 
$s_{k}^\infty$ and $s_{k}^2$ increase fast up to ${k}=15$ terms.
After that the rate of increase  is much smaller.
As expected, the small eigenvalues dominate the sum. 
Past $k=12$, the $L_\infty$ 
norm is stable while the $L_2$ norm continues to increase but at much
slower rate.

We did not carry out a full optimization of the amplitudes of the
generators since this is out of the scope of the article. Instead
we varied the amplitudes $G_i$ for to examine
how the power in the lines varies. We show two cases in the table below
\begin{table} [H]
\centering
\begin{tabular}{|c|c|c|c|c|c|c|c|c|}
   \hline
      & $G_1$ & $G_2$ & $G_{13}$ & $G_{22}$ & $G_{23}$ & $G_{27}$ & $ \parallel P_l \parallel _2$ & $ \parallel P_l \parallel _\infty$ \\ \hline
original& 23.54& 60.97& 37       & 21.59    & 19.2     & 26.91    &  68.78      &  37.\\ 
	case 2  & {\bf 3.54}& 60.97& 37       & 21.59    &{\bf 29.2}&{\bf 36.91}&{\bf 63.26} &{\bf 21.07}\\ 
	\hline
\end{tabular}
\caption{Two different configurations of generators for IEEE case 30 with their 
associated line powers $ \parallel P_l \parallel _2$ and $ \parallel P_l \parallel _\infty$. 
The terms that have changed from the 
original configuration are written in bold. }
\label{tab2}
\end{table}

\begin{figure}[H]
\centerline{\epsfig{file=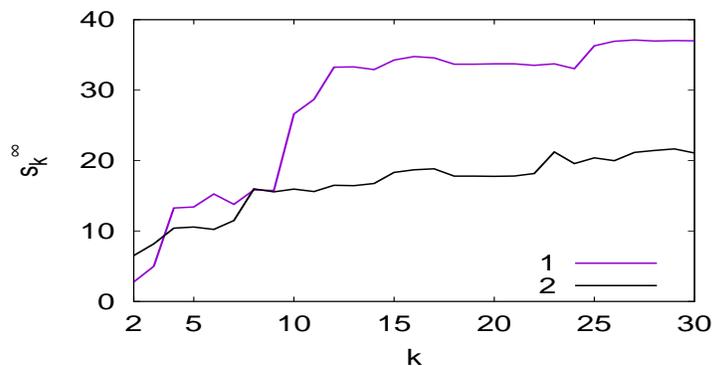,height=5cm,width=10cm,angle=0}}
\caption{
Plot of the partial sum (\ref{part_pli}) as a function of $n$
	for the two configurations original (0) and case 2, 
	shown in table \ref{tab2}.}
\label{part_pl}
\end{figure}

\begin{figure}[H]
\centerline{ \epsfig{file=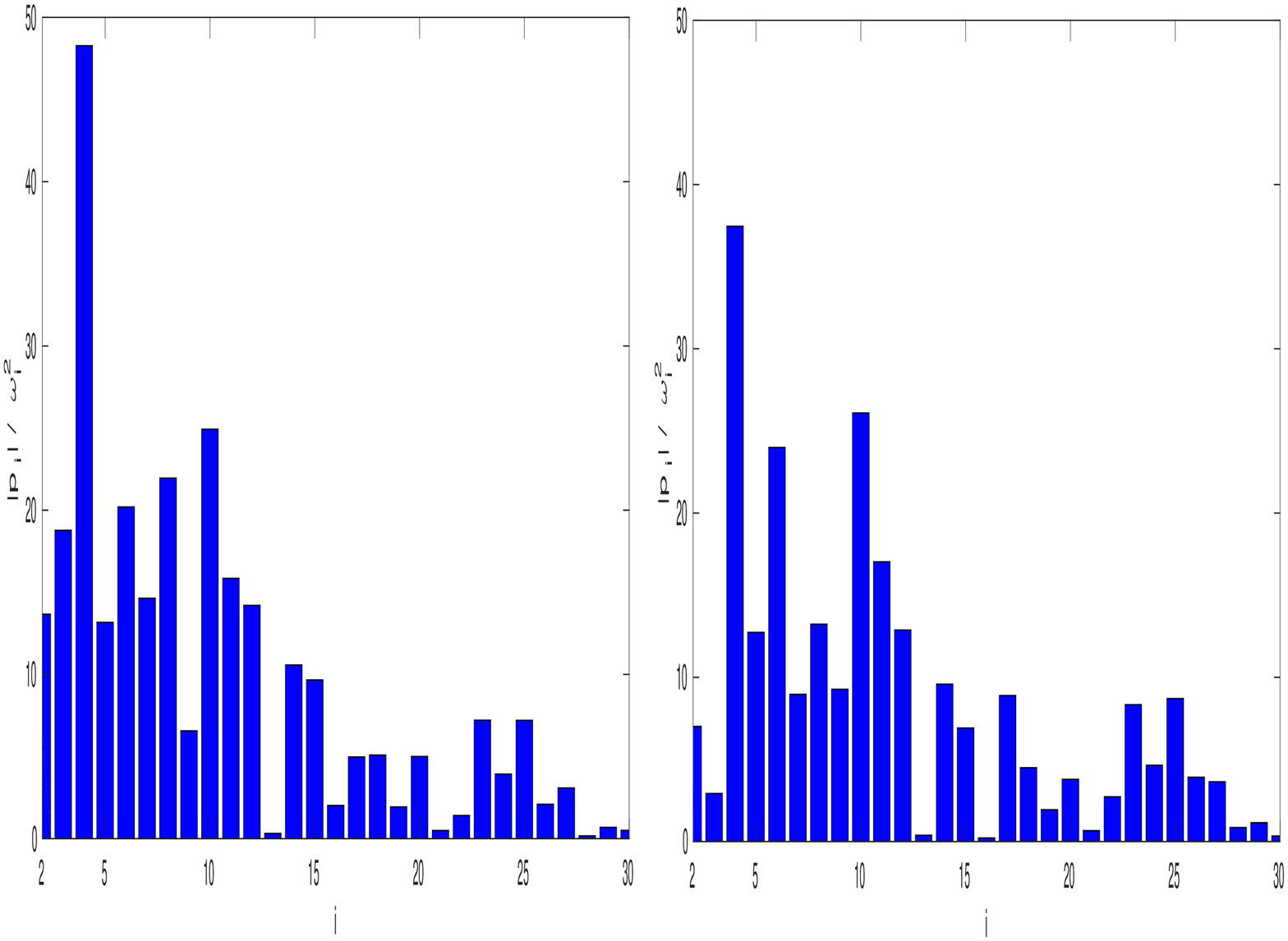,height=14cm,width=5cm,angle=90}}
\caption{
Plot of $|p_i| / \omega_i^2 $ for the original configuration (left) and the
improved configuration (right) shown in table \ref{tab2}.}
\label{ppo30}
\end{figure}

We computed the partial sum $s_{k}^\infty$ as a function of ${k}$
for the four different configurations of table \ref{tab2} in Fig. \ref{part_pl}.
The configuration 2 has a much lower value of $s_{k}^\infty$ than
the other configuration. To show the importance of the modal
distribution of power, we plot in Fig.  \ref{ppo30} $|p_i|/ \omega_i^2$ 
as a function of $i$ for the two configurations. 
Indeed, we see that configuration 2 has smaller $|p_i|$ for $i < 15$
than the original configuration. This explains the difference in 
$ \parallel P_l \parallel _2$ and especially $ \parallel P_l \parallel _\infty$.
This experiment shows that by tuning the amplitude of existing generators
one can decrease significantly the power in the lines.
We will carry out such an optimization in a further study.

\subsection{The IEEE 118 network} 
\begin{figure}[H]
\centerline{ \epsfig{file=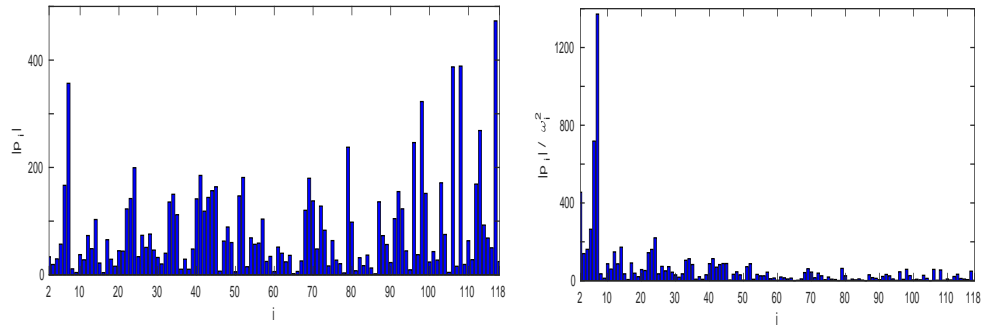,height=14cm,width=5cm,angle=90}}
\caption{
Plot of $|p_i|$ (left) and ${|p_i| \over \omega_i^2}$ (right) as a function of $i$ for the power vector $P$ of
IEEE case 118.}
\label{pp118}
\end{figure}

The components of the power vector are shown in Fig. \ref{pp118}.
A peak observed in $|p_7|$ in both panels. It corresponds to reinforcing the
localized eigenvector $\mathbf{v}^7$. The large components of $p_i$ are
smoothed out in the right panel by the denominator $\omega_i^2$.

\begin{figure}[H]
\centerline{\epsfig{file=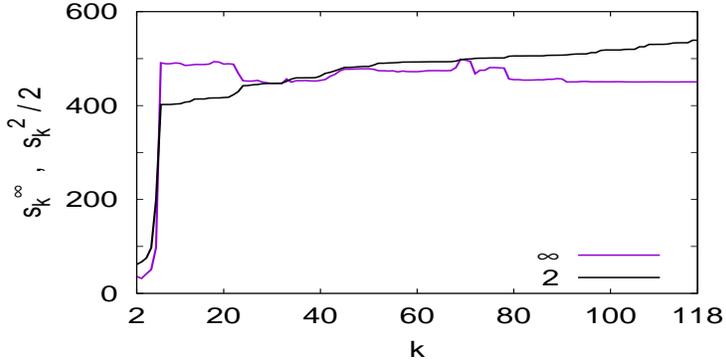,height=5cm,width=10cm,angle=0}}
\caption{
	Plot of the partial sums $s_{k}^\infty$ $s_{k}^2/2$ from
	(\ref{part_pli},\ref{part_pl2}) as a function of ${k}$.}
\label{pld2_118}
\end{figure}

We examine the convergence of $s_{k}^\infty, ~~s_{k}^2$ as ${k}$
increases. The graph is shown in Fig. \ref{pld2_118}.
As for case 30, both  $s_{k}^\infty$ and $s_{k}^2$ stabilize after 10 to 15 terms
and again the small eigenvalues dominate the sum.

\section{Conclusion and discussion}

We have shown that the load-flow equations can be reduced to a
singular linear system involving the graph Laplacian. Using the
a basis of eigenvectors of the Laplacian, we introduced
a spectral method to solve the load-flow equations. This provides a geometrical
picture of the power flow on the network, very similar to a Fourier
decomposition.

This spectral method provides an explicit expression of 
$ P_l$ as a sum of components $\nabla \mathbf{v}^i / \omega_i^2$,
where $\omega_i^2,~~ \mathbf{v}^i$ are respectively the $i$th eigenvalue 
and associated eigenvector of the Laplacian. These two components
play different roles. The eigenvalues $\omega_i^2$ typically increase with $i$
so that the small $i$ 's  will generally control the sum. The term
$\nabla \mathbf{v}^i$ is more difficult to estimate; it 
measures the space scale of the contribution on the network
and is loosely related to the nodal domains of $\mathbf{v}^i$.
Also, special eigenvectors $\mathbf{v}^i$ are strongly localized 
in a given region of the network and will dominate $P_l$ if $i$ is small.
Soft nodes, where the eigenvector has zero components also turned 
out to be important for optimization.

Using the orthogonality of $\mathbf{v}^i$, we obtained a 
Parseval-like expression of $ \parallel P_l \parallel _2$.
Numerical studies show 
that the main contribution to 
$ \parallel P_l \parallel _2$ and especially
to $ \parallel P_l \parallel _\infty$ tends to come from the small 
$i$ eigenvalues and eigenvectors, these correspond to 
large nodal domains i.e. large scales on the network. For example, 
only 10 or 20 modes are necessary to get a good estimate for a grid
network of 30 nodes. 
For a 118 node network, 15 modes are sufficient to describe the solution
with a 5 \% accuracy.  These numerical results are confirmed by
analysis done on a chain and a grid.

This geometric approach could complement the standard 
nonlinear load-flow because it gives a global
view of the network and the power vector.
Because of this, in view of the growing portion of intermittent sources,
our spectral approach could allow to optimize and reconfigure networks 
rapidly.

{\bf Acknowledgements}

The authors are funded by Agence Nationale de la Recherche grant "Fractal
grid". The calculations were done at the CRIANN computing center.


\begin{thebibliography}{99}
\bibitem{bc13} S. Backhaus and M. Chertkov, "Getting a grip on the electrical grid", Physics today 66 (5), 42 (2013).

\bibitem{kundur} P. Kundur, "Power System Stability and Control" ,
Mac Graw-Hill, (1994).

\bibitem{gsc15}  J. Grainger, Jr. W. Stevenson and Gary W. Chang ,
"Power Systems Analysis", McGraw-Hill  (2015).

\bibitem{crs01} D. Cvetkovic, P. Rowlinson and S. Simic, "An Introduction to the Theory of Graph Spectra",  London Mathematical Society Student Texts (No. 75), (2001).

\bibitem{ananum} G. Dahlquist, A. Bjorck and N. Anderson, "Numerical
methods", Prentice Hall, (1974).

\bibitem{panciatici} D. K. Molzahn, C. Josz, I. A. Hiskens and P. Panciatici, arxiv.1507.07212

\bibitem{cks13} J.G. Caputo, A. Knippel and E. Simo, J. Phys. A: Math. Theor. 46, 035100, (2013).

\bibitem{numrec}  W. H. Press , S. A. Teukolsky , W. T. Vetterling , B. P. Flannery , "Numerical Recipes: The Art of Scientific Computing", Cambridge University Press, (1986).
\bibitem{booklapla} T. Biyikoglu, J. Leydold and P. Stadler, "Laplacian eigenvectors of graphs", Springer (2000).
\bibitem{gladwell} E. B. Davies, G. M. L. Gladwell, J; Leydold and P. F. Stadler, "Discrete nodal domain theorems", Linear Algebra and its Applications 336 (2001) 51-60 .  
\bibitem{fiedler}  M. Fiedler, Algebraic connectivity of graphs, Czechoslovak Math. J., 23(98) (1973), 298-305.
\bibitem{mohar91} B. Mohar in "The Laplacian spectrum of graphs, Graph Theory, Combinatorics and Applications", Vol. 2, Ed. Y. Alavi, G.  Chartrand, O. R.  Oellermann, A. J. Schwenk, Wiley, pp. 871–898, (1991).  
\bibitem{matpower} http://www.pserc.cornell.edu/matpower/
\bibitem{case30} https://www2.ee.washington.edu/research/pstca/pf30/pg\_tca30bus.htm


\bibitem{graphviz} https://www.graphviz.org/

\end{thebibliography}
\end{document}